\documentclass[letterpaper,10pt,times,mathptm,psfig,conference]{IEEEtran}
\IEEEoverridecommandlockouts
\usepackage{cite}
\usepackage{amsmath,amssymb,amsfonts}
\usepackage{algorithmic}
\usepackage[ruled,linesnumbered]{algorithm2e}
\usepackage[normalem]{ulem}
\usepackage{amssymb}
\usepackage{amsmath}
\usepackage{graphicx}
\usepackage{textcomp}
\usepackage{hyperref}
\usepackage{booktabs}
\usepackage{xcolor}
\usepackage{multirow,array}
\usepackage{url}
\usepackage{epsfig, subfigure, amsmath, amssymb, wrapfig}
\def\BibTeX{{\rm B\kern-.05em{\sc i\kern-.025em b}\kern-.08em
    T\kern-.1667em\lower.7ex\hbox{E}\kern-.125emX}}

\newcommand{\khalil}[1]{}

\author{
Md Khalid Syfullah and Alvi Ataur Khalil\\
Computer Science, Southern Illinois University Carbondale, USA\\
\{mdkhalid.syfullah, a.khalil\}@siu.edu
\vspace{-15pt}
}

\begin{document}
\title{Can Quantum Layers offer Performance and Security Advantages in Vision-Language Models?}

\maketitle
\bstctlcite{IEEEexample:BSTcontrol}
\thispagestyle{empty}
\pagestyle{empty}

\begin{abstract}
As quantum computing moves from the noisy intermediate-scale quantum (NISQ) era toward early fault-tolerant quantum computing (EFTQC), quantum-enhanced AI is shifting from a conceptual idea to experimental research. Quantum components are expected to extend the capabilities of existing AI systems, yet it remains unclear how such components will help within large multimodal models. In this study, we try to answer that question for vision-language models (VLMs) using classically simulated quantum layers. With LLaVA-OneVision as the backbone, we evaluate 4- and 8-qubit basic-ring and strongly entangling circuits at four insertion and four replacement locations. Insertion adds quantum layers to the classical model, while replacement substitutes them for classical components. Using the same setup, we compare performance and computational cost between classical and quantum-classical hybrid models. On the Visual Spatial Reasoning dataset, inserting a quantum layer after the visual projector reaches 80.7\% development accuracy, 2.28 percentage-points above classical full fine-tuning. Replacing the final decoder feed-forward sublayer retains 79.22\% accuracy with 13.07 million fewer trainable parameters. Quantum insertion can thus improve accuracy, and selective replacement can reduce parameters. These benefits depend strongly on where the quantum layer is inserted or which classical component it replaces: earlier replacements lower accuracy, and fewer parameters do not reliably shorten training or reduce computation. To assess potential security benefits, we evaluated the robustness of quantum-classical hybrid models under six attack settings. Overall, our study maps where quantum layers help and hurt within VLMs, offering a starting point for hybrid VLM design; whether these gains generalize remains open. The code is available at \url{https://github.com/TITANS-Lab-Artifacts/QuantumVLM}.
\end{abstract}

\begin{IEEEkeywords}
Quantum machine learning, vision-language models, adversarial robustness, variational quantum circuits
\end{IEEEkeywords}

\section{Introduction}
\label{sec:introduction}
Vision-language models (VLMs) are widely used for tasks involving images and text~\cite{radford2021learning}. One such task is Visual Question Answering (VQA), answering natural-language questions about an image, with applications in education, finance, and healthcare. The privacy-sensitive and high-stakes nature of these domains requires VLMs to be highly reliable, accurate, and robust. Developing VLMs with these capabilities requires significant computational resources, as VQA is often considered an AI-complete problem~\cite{antol2015vqa}. Despite this complexity, VLMs remain vulnerable to attack families including privacy breaches, model extraction, data extraction, and reverse engineering~\cite{liu2024surveyattacks}. These issues motivate research into making VLMs more cost-effective and robust.

Quantum machine learning (QML) is one possible direction for addressing complex learning problems, with the potential for quantum advantage~\cite{biamonte2017quantum}. Till now, early fault-tolerant quantum systems remain constrained by residual noise, limited logical qubits, and restricted circuit depth~\cite{preskill2018nisq,palsberg2026early}. Outputs from current machines contain noise, making practical quantum advantage difficult to achieve~\cite{cerezo2022challenges}. However, quantum simulators can simulate ideal circuits using classical resources~\cite{suzuki2021qulacs}. A full state-vector simulation of $n$ qubits stores $2^n$ complex amplitudes, each requiring 16 bytes in double precision~\cite{qiskitaer_statevector}.

Despite these hardware limitations, the prospect of quantum-enhanced AI has motivated research into quantum methods for improving machine learning~\cite{biamonte2017quantum}. For language-model fine-tuning, quantum parameter adaptation (QPA) has reduced trainable parameters in methods such as low-rank adaptation (LoRA) while maintaining comparable or better performance~\cite{liu2025qpa}. Quantum layers have also been combined with vision models for image classification~\cite{mari2020transfer}, and quantum circuits have been applied to language-processing tasks~\cite{meichanetzidis2021language}. These works motivate investigating whether quantum layers can improve VLM performance. Experimental quantum adversarial learning has also demonstrated vulnerabilities in quantum classifiers and explored adversarial training as a defense~\cite{ren2022adversarial}. This motivates testing whether quantum integration improves VLM robustness rather than assuming such benefits.

In this work, we use classical simulations of quantum layers to investigate how they affect three aspects of a VLM: (i) predictive accuracy, (ii) computational cost, and (iii) adversarial robustness. We use the Large Language and Vision Assistant (LLaVA) OneVision model, with its Qwen2-0.5B single-image (SI) variant as the backbone~\cite{llavahf2024model}. We examine four quantum circuits: 4- and 8-qubit basic-ring (BR4, BR8) and strongly entangling (SE4, SE8) circuits. We test two approaches: insertion and replacement. For insertion, we add quantum layers at four positions within the VLM: (i) before the visual projector, (ii) after the visual projector, (iii) after an early decoder block, and (iv) after the final decoder normalization. These positions let us examine quantum processing of visual features, projected visual features, early multimodal representations, and final representations used for prediction, respectively. For replacement, we substitute quantum layers for the visual-to-language projector or the feed-forward sublayers in early, intermediate, and final decoder blocks. We evaluate six untargeted adversarial image-perturbation attacks, grouped by attacker access: (i) the fast gradient sign method (FGSM)~\cite{goodfellow2015adversarial}, (ii) projected gradient descent with a cross-entropy objective (PGD-CE), and (iii) PGD with a margin objective (PGD-margin), all using target-model gradients~\cite{madry2018robust,carlini2017evaluating}; (iv) simultaneous perturbation stochastic approximation (SPSA), using target-model scores~\cite{uesato2018risk}; (v) HopSkipJump (HSJ), using predicted labels only~\cite{chen2020hopskipjump}; and (vi) transfer PGD (T-PGD), using a surrogate model~\cite{papernot2017practical}. Each attack modifies the image to cause an incorrect prediction while keeping the text prompt and weights fixed. Our contributions are threefold:

\begin{itemize}
\item We train and evaluate quantum-layer configurations and classical baselines on the same dataset to assess how circuit design and placement affect predictive accuracy.

\item We compare trainable parameters, training time, and memory usage to assess whether any accuracy gains justify the computational cost of quantum simulation.

\item We test whether quantum layers improve robustness against these six attacks using a common protocol.
\end{itemize}

These experiments address three research questions (RQs):
\begin{itemize}
\item \textbf{RQ1:} How does inserting quantum layers at different locations within a VLM affect predictive performance and computational cost?
\item \textbf{RQ2:} Can quantum layers effectively replace classical components while maintaining predictive performance and reducing computational cost?
\item \textbf{RQ3:} How do VLMs with quantum-layer insertion or replacement compare with classical baselines in adversarial robustness under different levels of attacker access?
\end{itemize}

Paper Organization: Section~\ref{sec:background} presents the background. Section~\ref{sec:literature_review} discusses related work. Section~\ref{sec:proposed_framework} presents the framework. Section~\ref{sec:methodology} outlines the methodology. Section~\ref{sec:experimental-results} reports the results, and Section~\ref{sec:conclusion} concludes the paper.

\section{Background}
\label{sec:background}
This section introduces the concepts underlying our work.

\subsection{VLM Components}
\label{subsec:vlm_components}
In LLaVA-OneVision~\cite{li2025onevision}, the vision encoder converts an image into features representing its visual content. The projector converts these features into the format expected by the language model. The text prompt is converted into token embeddings and combined with the projected image features. Decoder blocks then process this combined sequence: attention connects information across tokens, helping relate the text to the image, while feed-forward sublayers transform each token's features. Normalization controls feature scales, and residual connections carry information across processing stages. Finally, the output head maps normalized decoder features to token scores for True or False prediction.

\subsection{Attacks on VLMs}
\label{subsec:vlm_attacks}
\begin{figure}[t]
    \centering
    \includegraphics[width=\columnwidth]{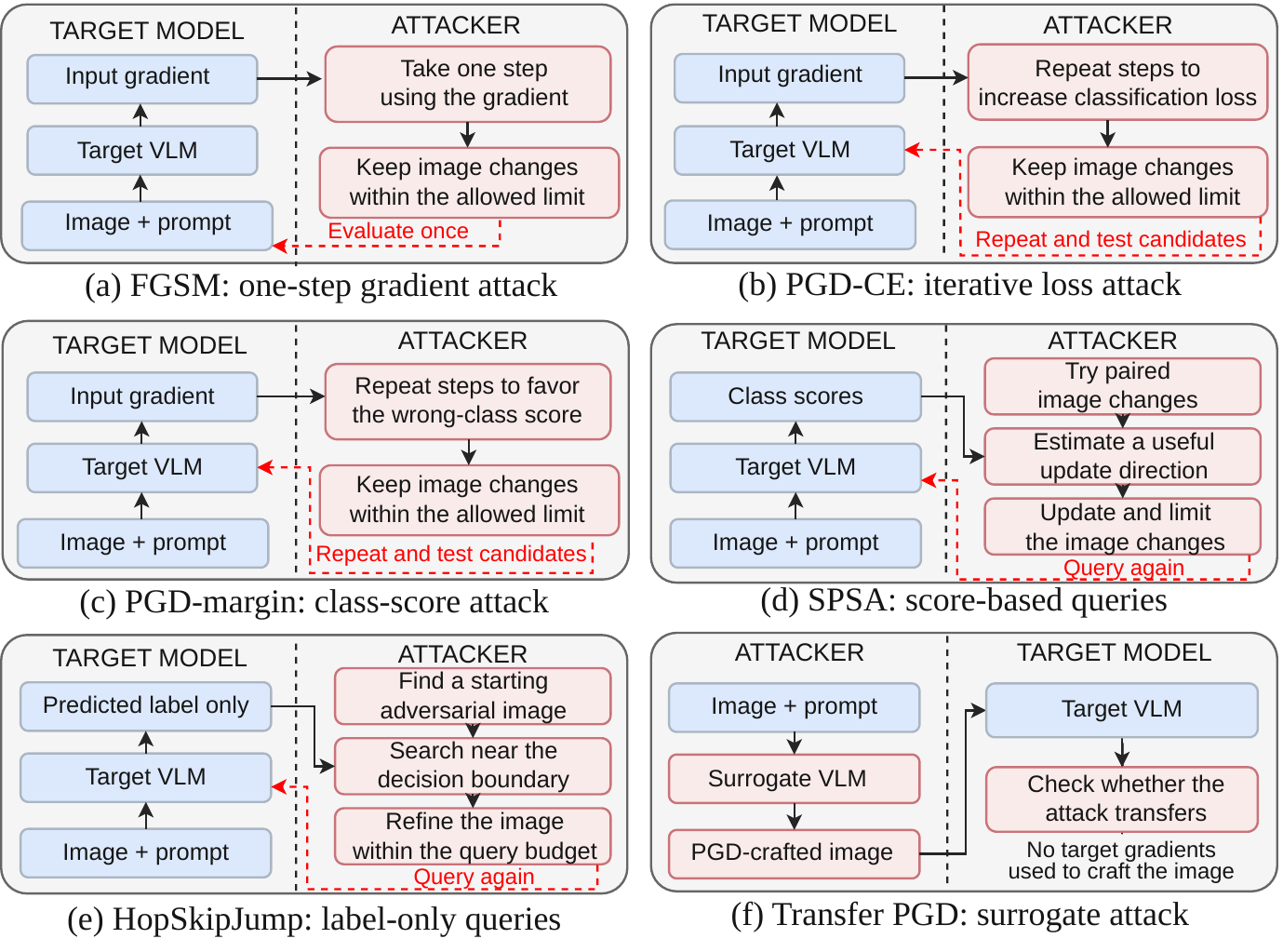}
    \vspace{-20pt}
    \caption{Overview and access level of the attack methods used in this work.}
    \vspace{-15pt}
    \label{fig:attack_overview}
\end{figure}

We compare classical and quantum VLMs under six attacks that modify image pixels, with different levels of attacker access, shown in Fig.~\ref{fig:attack_overview}. FGSM uses the target model's gradients to increase prediction loss in one step~\cite{goodfellow2015adversarial}. PGD-CE and PGD-margin repeatedly modify the image within a fixed limit: PGD-CE increases cross-entropy loss, while PGD-margin raises the incorrect class score relative to the correct one~\cite{madry2018robust,carlini2017evaluating}. SPSA estimates how to modify the image from output scores without target gradients~\cite{uesato2018risk}. HSJ uses predicted labels to make an adversarial image closer to the original~\cite{chen2020hopskipjump}. T-PGD creates perturbations on a separate surrogate model and tests whether they also fool the target VLM~\cite{papernot2017practical}.

\section{Literature Review}
\label{sec:literature_review}

Quantum approaches to vision and language tasks target model architecture or parameter-efficient adaptation. Cherrat et al.~\cite{cherrat2024quantumvision} introduced quantum vision transformers with quantum attention, and Unlu et al.~\cite{unlu2024hybrid} reported competitive classification performance with much longer training than classical models. Liu et al.~\cite{liu2025qpa} generated language-model adaptation parameters with quantum circuits and a classical mapping, keeping inference classical. Chen et al.~\cite{chen2024quanta} proposed QuanTA, a quantum-inspired classical method for high-rank fine-tuning. Kong et al.~\cite{kong2025hyqut} replaced parts of language transformers with variational quantum circuits and reported comparable generation performance. These studies show promise for quantum architectures and compact adaptation; none tests whether they reduce overall computational cost in VLMs.

Several studies have applied quantum methods to multimodal learning. Mukesh et al.~\cite{mukesh2024qvila} combined a classical vision-language transformer with quantum image-text fusion. Zhao et al.~\cite{zhao2026mqadapter} proposed a quantum adapter using category-based semantic anchors to refine visual features for few-shot VLM adaptation. Wang and Shang~\cite{wang2026pqfa} applied parallel quantum circuits after image-text fusion, comparing against a classical MLP and testing missing modalities and simulated circuit noise. These studies demonstrate quantum adaptation and fusion without testing quantum layers replacing classical layers in the same VLM under common adversarial attacks.

Adversarial robustness complements accuracy and computational cost when evaluating quantum layers. Ren et al.~\cite{ren2022adversarial} demonstrated adversarial vulnerability in quantum classifiers on superconducting hardware and tested adversarial training as a defense. Athalye et al.~\cite{athalye2018obfuscated} showed that obfuscated gradients can falsely suggest robustness, motivating evaluation with gradient-based, query-based, and transfer attacks. These findings call for direct adversarial evaluation; how quantum circuit design and placement affect robustness within the same VLM remains unclear.

Prior studies leave three gaps: how inserting quantum layers, or replacing classical layers with them, affects VLM performance; whether accuracy gains justify computational cost; and how quantum-enabled VLMs perform under adversarial attacks. Using a shared LLaVA-OneVision backbone, we test four quantum circuits at four insertion and four replacement locations. We compare accuracy, trainable parameters, training time, and memory with classical baselines, then evaluate six attacks with shared perturbation budgets and gradient, score, label, or surrogate-model access. We jointly examine how circuit design and placement affect accuracy, computational cost, and adversarial robustness.

\section{Proposed Framework}
\label{sec:proposed_framework}

\begin{figure}[!t]
    \centering
    \includegraphics[width=\columnwidth]{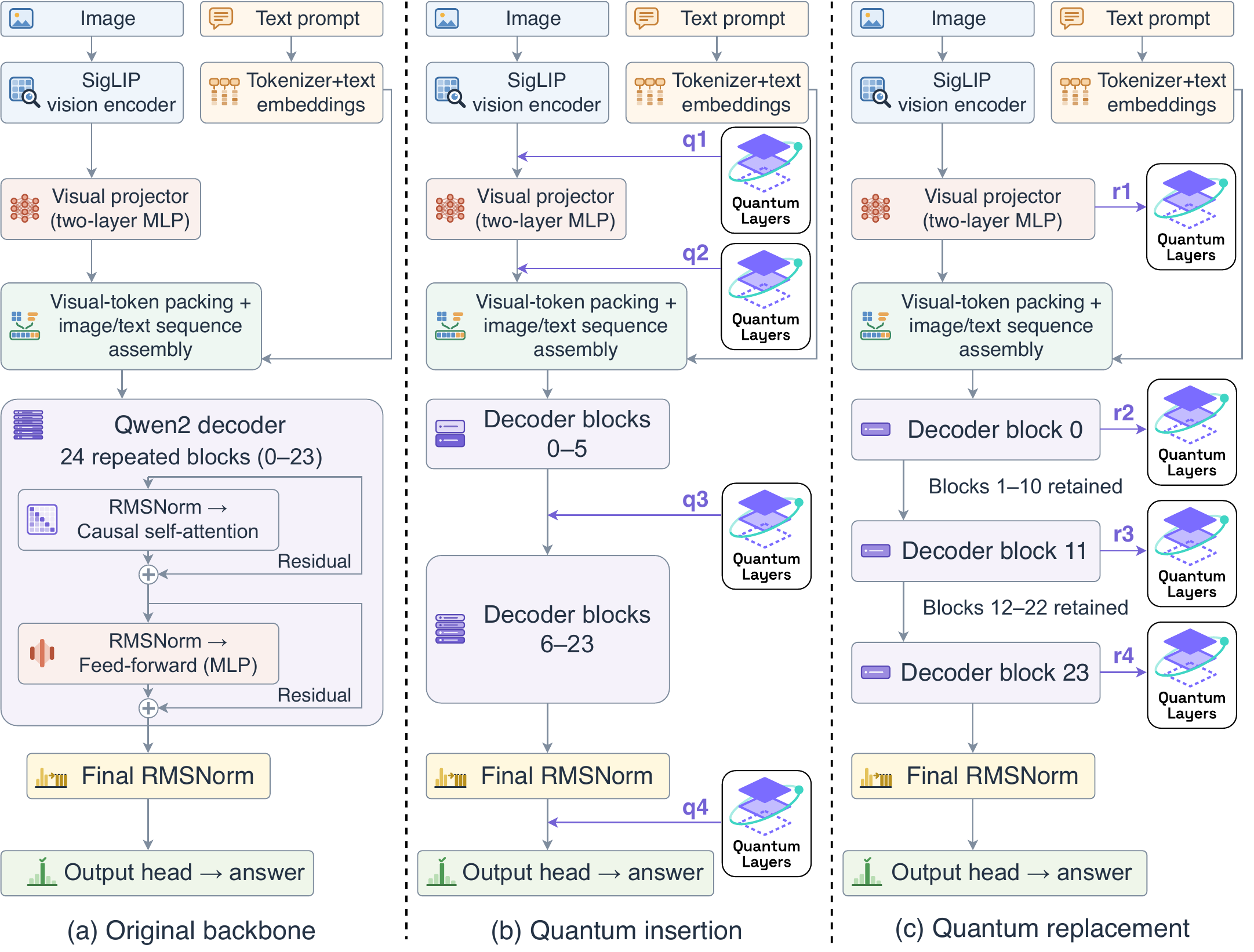}
    \vspace{-20pt}
    \caption{Proposed framework showing the LLaVA-OneVision-Qwen2-0.5B-SI backbone and quantum-layer insertion and replacement locations.}
    \vspace{-20pt}
    \label{fig:architecture}
\end{figure}

Fig.~\ref{fig:architecture} shows the proposed framework for quantum insertion and replacement in backbone VLM. The vision encoder uses Sigmoid Loss for Language-Image Pre-training (SigLIP), and the decoder uses root mean square normalization (RMSNorm).

\subsection{Quantum Insertion Locations}
Insertion adds a quantum layer to the model without removing any existing component.

\begin{itemize}
    
\item \textbf{Before the visual projector (q1):} We place the quantum layer on vision-encoder features before they reach the projector, to test whether adjusting image features before entering model's feature space improves prediction.

\item \textbf{After the visual projector (q2):} We insert the quantum layer after the projector, before image features are combined with text embeddings. It tests whether this adjustment helps the model to fuse image and text information.

\item \textbf{After an early decoder block (q3):} We apply the quantum layer to the final prompt token after decoder block 5, adjusting image-text information early in decoding improves prediction.

\item \textbf{After final decoder normalization (q4):} We position the quantum layer between the final normalization and the output head, acting on the final prompt token, to test whether adjusting the fully processed representation just before prediction improves accuracy.
\end{itemize}



\subsection{Quantum Replacement Positions}

Unlike insertion, replacement swaps one classical component for a quantum module. We use linear input and output projections around the module to ensure that it receives and returns features of the same dimension as the original component. Thus, the rest of the model is unchanged. When replacing a decoder MLP, only the MLP is swapped; the block's attention, normalization, and residual connections remain in place. 

\begin{itemize}
\item \textbf{Complete visual projector (r1):} Here, we replace the entire projector with a quantum module. This tests whether visual features can be mapped into the language model's feature space.

\item \textbf{Early decoder MLP (r2):} We substitute a quantum module for the MLP in decoder block 0 to examine whether it can transform features at the start of decoding.

\item \textbf{Intermediate decoder MLP (r3):} We swap the MLP in decoder block 11 for a quantum module. This assesses whether it can transform features after several stages of image-text interaction.

\item \textbf{Final decoder MLP (r4):} We replace the MLP in decoder block 23 with a quantum module to determine whether it can perform the final feature transformation before normalization and prediction.
\end{itemize}

\begin{table}[t!]
\centering
\caption{Summary of notation.}
\vspace{-5pt}
\label{tab:notation}
\setlength{\tabcolsep}{3pt}
\renewcommand{\arraystretch}{1.05}
\scriptsize
\begin{tabular}{|>{\centering\arraybackslash}p{0.18\columnwidth}|>{\raggedright\arraybackslash}p{0.74\columnwidth}|}
\hline
\textbf{Symbol} & \textbf{Meaning} \\
\hline
$x,t,y$ & Image, statement, and label $y\in\{0,1\}$ \\
\hline
$x^{\mathrm{adv}},\delta,\epsilon$ & Adversarial image, perturbation, and maximum pixel change \\
\hline
$p_y$ & Probability for class $y$ \\
\hline
$h,h',d,d_{\mathrm{out}}$ & Input and updated features, input and output dimensions \\
\hline
$n_q,L,j$ & Circuit qubit count, depth, and qubit index \\
\hline
$W_{\mathrm{in}},b_{\mathrm{in}}$ & Input projection weights and bias \\
\hline
$W_{\mathrm{out}},b_{\mathrm{out}}$ & Output projection weights and bias \\
\hline
$\mathcal N,z,\theta$ & Normalization, input angles, and circuit parameters \\
\hline
$V,U_\theta,\lvert\psi\rangle$ & Encoding circuit, trainable circuit, and quantum state \\
\hline
$Z_j,m$ & Pauli-Z operator on qubit $j$ and measurement vector \\
\hline
$g_\theta,\gamma$ & Quantum module output and trainable residual scale \\
\hline
$\mathcal L_{\mathrm{CE}}$ & CE loss \\
\hline
$\alpha$ & Attack step size \\
\hline
$N,N_c,N_s$ & Total examples, initially correct examples, and successfully attacked examples \\
\hline
$N_k,C_k$ & Total and correctly classified examples in class $k\in\{0,1\}$ \\
\hline
$\|\cdot\|_\infty,\mathbb R^d$ & Maximum absolute entry and $d$-dimensional real space \\
\hline
\end{tabular}
\vspace{-10pt}
\end{table}

\section{Methodology}
\label{sec:methodology}

This section describes our technical framework, details of used quantum circuits, training and attack setup, and the metrics used for evaluation. Table~\ref{tab:notation} defines the notation.

\subsection{Quantum Circuits}
\begin{figure}[!t]
    \centering
    \includegraphics[width=\columnwidth]{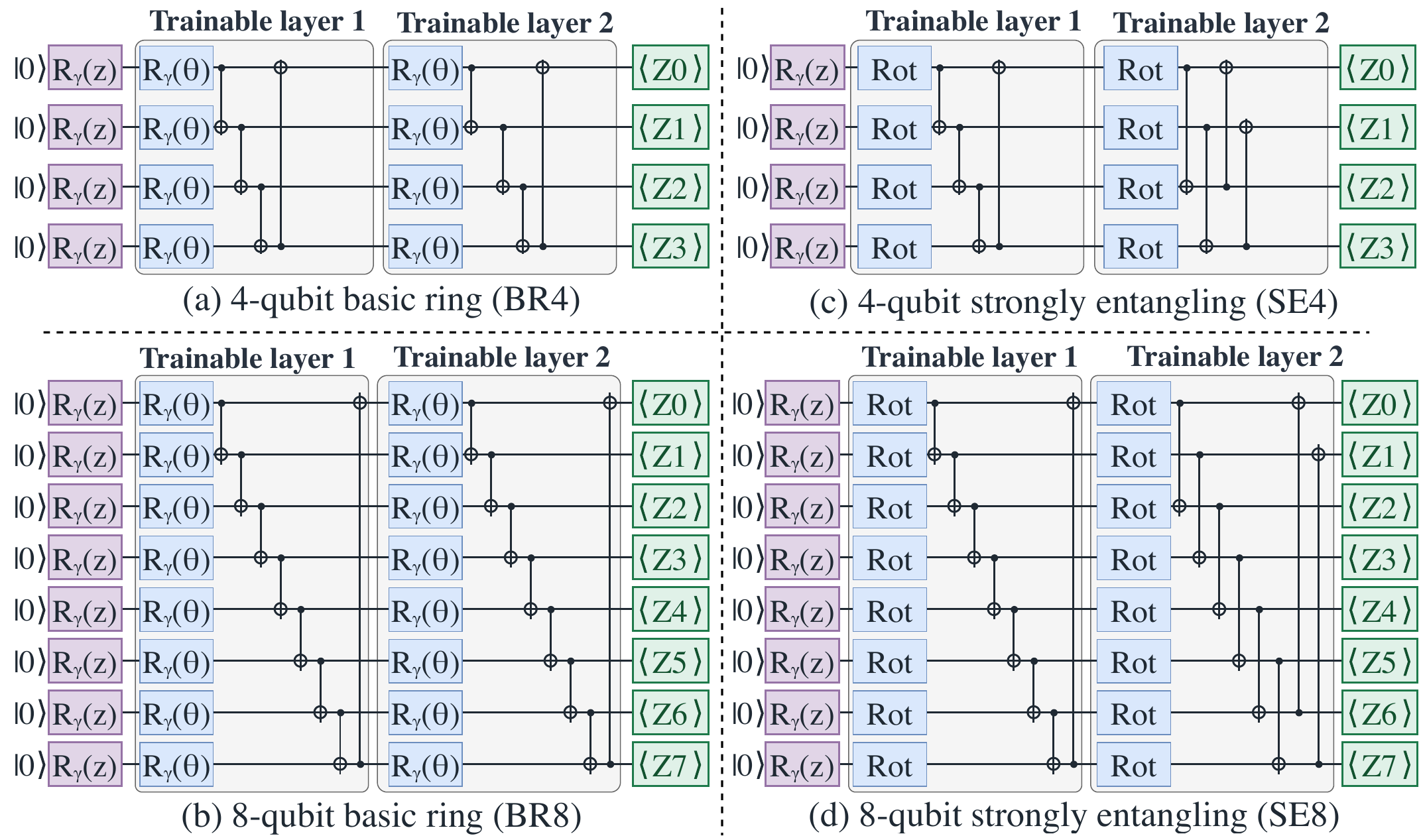}
    \vspace{-20pt}
    \caption{Four quantum circuits used in this study.}
    \vspace{-15pt}
    \label{fig:quantum_circuits}
\end{figure}

We test two circuit families, each with 4 and 8 qubits, giving the four configurations shown in Fig.~\ref{fig:quantum_circuits}. The basic-ring circuit is the simpler design: each layer applies a trainable Y-axis rotation to every qubit, then links neighboring qubits in a closed ring with controlled-NOT (CNOT) gates. The strongly entangling circuit is more expressive: each layer applies a trainable Rot gate, a sequence of Z-, Y-, and Z-axis rotations, to every qubit, then uses CNOT gates whose connection range changes from layer to layer. 

\subsection{Technical Framework}

We use the same image preprocessing, prompt, and binary task for classical and quantum models to isolate differences arising from the quantum component. Given an image $x$ and a statement $t$, the model predicts True or False, trained with $\mathcal L_{\mathrm{CE}}=-\log p_y$, where $p_y$ is the probability of the correct label $y$. For each selected token feature $h\in\mathbb R^d$, we map the normalized feature to $n_q$ rotation angles $z=\pi\tanh(W_{\mathrm{in}}\mathcal N(h)+b_{\mathrm{in}})$. We encode $z$ once with Y-axis rotations $V(z)$, apply $L$ trainable layers $U_\theta$ to obtain $\lvert\psi\rangle=U_\theta V(z)\lvert0\rangle^{\otimes n_q}$, and measure each qubit's Pauli-Z expectation $m_j=\langle\psi\rvert Z_j\lvert\psi\rangle$.

An output projection gives $g_\theta(h)=W_{\mathrm{out}}m+b_{\mathrm{out}}$. For insertion, we use $h'=h+\gamma g_\theta(h)$, with trainable $\gamma$ initialized to 0.1 to limit early changes to pretrained features. For replacement, the $d_{\mathrm{out}}$-dimensional $g_\theta(h)$ is the component's output; the module adds no residual, decoder residuals remain.

Insertion updates all retained visual tokens at q1 and q2, and only the final prompt token at q3 and q4. Replacement processes all visual tokens at r1 and all tokens entering the selected MLP at r2--r4. Insertion and projector replacement use layer normalization without learnable scale and shift; decoder replacements reuse the decoder's normalized features. We simulate circuits exactly with a differentiable PyTorch state-vector simulator, without shot sampling or hardware noise.


\subsection{Attack Setup}

We attack only full-backbone models, where the quantum module and classical layers are trained together, to evaluate the robustness of the complete hybrid model. Attacks perturb the image before preprocessing, while the text input and model weights stay fixed. An attack succeeds if the model makes an incorrect prediction on an image that satisfies Eq.~\eqref{eq:perturbation}:

\vspace{-10pt}
\begin{equation}
\label{eq:perturbation}
x^{\mathrm{adv}}=x+\delta,\qquad
\|\delta\|_\infty\leq\epsilon,\qquad
x^{\mathrm{adv}}\in[0,1].
\end{equation}

We run all attacks on the first 40 development examples with $\epsilon=4/255$. The attacks cover three levels of attacker knowledge: gradient access (FGSM, PGD), score or label access only (SPSA, HSJ), and no access to the target model (transfer PGD). The attack setups are:

\paragraph{FGSM~\cite{goodfellow2015adversarial}} A single step of size $4/255$ that increases the CE loss, starting from the original image.

\paragraph{PGD-CE and PGD-margin~\cite{madry2018robust,carlini2017evaluating}} Five random starts with 100 updates each and step size $\alpha=2.5\epsilon/100$, projected back into the allowed perturbation after every update. PGD-CE increases the CE loss; PGD-margin increases the incorrect class score relative to the correct one.

\paragraph{SPSA~\cite{uesato2018risk}} Estimates gradients from model scores alone. Each of at most 100 updates compares 16 pairs of opposite perturbations of radius 0.01, followed by a projected Adam step of size $\epsilon/10$; we use one restart.

\paragraph{HSJ~\cite{chen2020hopskipjump}} Uses only predicted labels. Starting from a misclassified image, up to 32 iterations shrink the perturbation while keeping the prediction incorrect.

\paragraph{T-PGD~\cite{papernot2017practical}} We run PGD-CE with same settings on a separate surrogate VLM, with a frozen backbone and only its output head trained, and test the resulting images.


\subsection{Evaluation Metrics}

We report clean accuracy, balanced accuracy (BA), true recall (TR), and false recall (FR); for each attack and their union, we also report robust accuracy (RA). BA weights both classes equally, and both BA and RA are defined in Eq.~\eqref{eq:metrics}:
\begin{equation}
\label{eq:metrics}
\mathrm{BA}=\frac{1}{2}\left(\frac{C_0}{N_0}+\frac{C_1}{N_1}\right),\qquad
\mathrm{RA}=\frac{N_c-N_s}{N}.
\end{equation}

Confidence intervals use 1,000 bootstrap resamples grouped by image. For resources, we compare trainable parameters (params) in millions (M), training time, and peak memory.

\subsection{Training Setup}

We use the random split of the Visual Spatial Reasoning (VSR) dataset~\cite{liu2023vsr}, with 7,680 training and 1,097 development examples. Classical baselines are zero-shot evaluation, head-only training, projector-only training, LoRA, and full fine-tuning. Insertion trains the quantum module alone or with the head, projector, LoRA adapters, or full backbone; replacement uses full-backbone training. Each run uses one NVIDIA RTX PRO 6000 Blackwell GPU (96~GB), seed 0, up to 20 epochs with early stopping (patience 2--5), and keeps the checkpoint with the lowest development loss. We use AdamW~\cite{loshchilov2019decoupled} with weight decay 0.01, gradient clipping at 1.0, 5\% warm-up, linear decay, and an effective batch size of 16. The learning rate is $10^{-5}$ for projector and full-backbone training and $10^{-4}$ otherwise. LoRA~\cite{hu2022lora} uses rank 8, scaling 16, and dropout 0.05 on the decoder query and value projections. 


\begin{table}[t]
\centering
\caption{Performance of classical baselines.}
\vspace{-5pt}
\label{tab:table_1_classical}
\begingroup
\scriptsize
\setlength{\tabcolsep}{1.5pt}
\renewcommand{\arraystretch}{0.9}
\begin{tabular*}{\columnwidth}{@{\extracolsep{\fill}}lccccc@{}}
\hline
Mode &
\begin{tabular}[c]{@{}c@{}}Params(M)\end{tabular} &
Accuracy (\%) &
\begin{tabular}[c]{@{}c@{}}BA (\%)\end{tabular} &
\begin{tabular}[c]{@{}c@{}}TR (\%) \end{tabular} &
\begin{tabular}[c]{@{}c@{}}FR (\%)\end{tabular} \\
\hline
Zero-shot/Full-Frozen        & 0.000   & 54.2 & 53.1 & 92.4 & 13.9 \\
Head only        & 0.002   & 64.9 & 65.0 & 62.6 & 67.4 \\
Projector        & 1.837   & 58.5 & 58.4 & 62.8 & 54.0 \\
LoRA             & 0.541   & 79.0 & 79.0 & 79.3 & 78.8 \\
Full fine-tuning & 893.676 & 78.4 & 78.3 & 81.6 & 75.0 \\
\hline
\end{tabular*}
\vspace{-15pt}
\endgroup
\end{table}

\section{Experimental Results}
\label{sec:experimental-results}

This section compares classical VLM baselines with quantum insertion and replacement models in predictive performance, computational cost, and adversarial robustness.

\subsection{Performance of Classical Baselines}

Table~\ref{tab:table_1_classical} compares performance across different classical model configurations. Zero-shot and fully frozen evaluation achieve 54.2\% accuracy, with True and False recall of 92.4\% and 13.9\%, showing a strong bias toward True. LoRA achieves the highest classical accuracy at 79.0\%, compared with 78.4\% for full fine-tuning, while training approximately 0.54 million parameters rather than 893.68 million.

\subsection{Performance of Quantum Insertion and Replacement}

\begin{figure}[t]
\centering
    \includegraphics[width=\columnwidth]{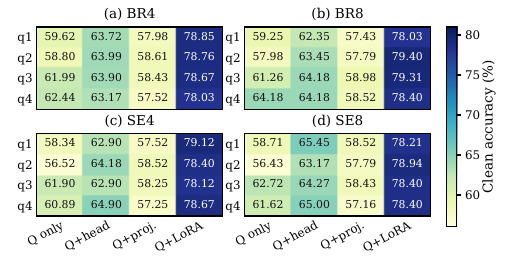}
    \vspace{-30pt}
    \caption{Accuracy (\%) across quantum insertion training modes.}
    \vspace{-10pt}
\label{fig:results_1}
\end{figure}

\begin{figure}[t]
\centering
    \includegraphics[width=\columnwidth]{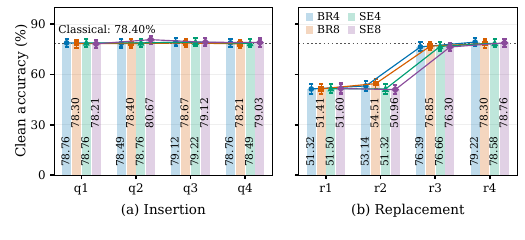}
    \vspace{-30pt}
    \caption{Accuracy (\%) under full-backbone training. Error bars show 95\% image-cluster bootstrap intervals. Dotted lines mark classical full fine-tuning.}
    \vspace{-10pt}
\label{fig:results_2}
\end{figure}

Figs.~\ref{fig:results_1} and~\ref{fig:results_2} show that quantum insertion depends on which components are trained. Training only the quantum module gives 56.4 to 64.2\% accuracy, while joint training with LoRA or the full backbone reaches approximately 78 to 81\%. Under full fine-tuning, SE8 at q2 achieves the highest accuracy and balanced accuracy, both 80.7\% (Table~\ref{tab:full_training}). Its unrounded accuracy exceeds classical full fine-tuning by 2.28 percentage points. Increasing the qubit count or changing the circuit does not consistently improve accuracy at other positions.

Fig.~\ref{fig:results_2}(b) shows that replacing the visual projector at r1 or the early feed-forward sublayer at r2 reduces accuracy to approximately 51 to 55\%. Intermediate replacement at r3 retains approximately 76 to 77\%, while final replacement at r4 remains close to the classical baseline. BR4 at r4 achieves 79.22\% accuracy with approximately 880.61 million trainable parameters, removing approximately 13.07 million parameters while preserving accuracy in this run.

\begin{table}[!t]
\centering
\caption{Performance under full fine-tuning.}
\vspace{-5pt}
\label{tab:full_training}
\begingroup
\scriptsize
\setlength{\tabcolsep}{1pt}
\renewcommand{\arraystretch}{0.95}
\begin{tabular*}{\columnwidth}{@{\extracolsep{\fill}}ccccc@{}c@{}|cccc@{}}
\hline
\multirow[c]{2}{*}{Circuit} & \multicolumn{4}{c}{Insertion} & & \multicolumn{4}{c}{Replacement} \\
\cline{2-5}\cline{7-10}
 & Pos. & Accuracy(\%) &
\begin{tabular}[c]{@{}c@{}}BA(\%)\end{tabular} &
\begin{tabular}[c]{@{}c@{}}Params(M)\end{tabular} & &
Pos. & Accuracy(\%) &
\begin{tabular}[c]{@{}c@{}}BA(\%)\end{tabular} &
\begin{tabular}[c]{@{}c@{}}Params(M)\end{tabular} \\
\hline
\multirow[c]{4}{*}{BR4} & q1 & 78.8 & 78.7 & 893.69 & & r1 & 51.3 & 50.0 & 891.85 \\
 & q2 & 78.5 & 78.3 & 893.68 & & r2 & 53.1 & 52.9 & 880.61 \\
 & q3 & 79.1 & 79.1 & 893.68 & & r3 & 76.4 & 76.0 & 880.61 \\
 & q4 & 78.8 & 78.7 & 893.68 & & r4 & 79.2 & 79.1 & 880.61 \\
\hline
\multirow[c]{4}{*}{BR8} & q1 & 78.3 & 78.2 & 893.70 & & r1 & 51.4 & 50.0 & 891.86 \\
 & q2 & 78.4 & 78.3 & 893.69 & & r2 & 54.5 & 54.4 & 880.62 \\
 & q3 & 78.7 & 78.7 & 893.69 & & r3 & 76.8 & 76.7 & 880.62 \\
 & q4 & 78.2 & 78.1 & 893.69 & & r4 & 78.3 & 78.2 & 880.62 \\
\hline
\multirow[c]{4}{*}{SE4} & q1 & 78.8 & 78.7 & 893.69 & & r1 & 51.5 & 51.2 & 891.85 \\
 & q2 & 78.8 & 78.7 & 893.68 & & r2 & 51.3 & 50.5 & 880.61 \\
 & q3 & 79.2 & 79.1 & 893.68 & & r3 & 76.7 & 76.6 & 880.61 \\
 & q4 & 78.5 & 78.4 & 893.68 & & r4 & 78.6 & 78.5 & 880.61 \\
\hline
\multirow[c]{4}{*}{SE8} & q1 & 78.2 & 78.2 & 893.70 & & r1 & 51.6 & 50.2 & 891.86 \\
 & q2 & 80.7 & 80.7 & 893.69 & & r2 & 51.0 & 49.9 & 880.62 \\
 & q3 & 79.1 & 79.1 & 893.69 & & r3 & 76.3 & 76.1 & 880.62 \\
 & q4 & 79.0 & 78.9 & 893.69 & & r4 & 78.8 & 78.6 & 880.62 \\
\hline
\end{tabular*}
\endgroup
\vspace{-10pt}
\end{table}

\subsection{Computational Cost}
\begin{figure}[!t]
\centering
    \includegraphics[width=\columnwidth]{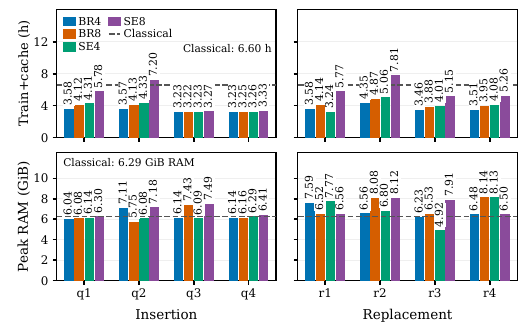}
    \vspace{-20pt}
    \caption{Training time (top) and peak RAM usage (bottom) under full-backbone training. Dashed lines show the classical baseline. Only training is measured.}
    \vspace{-15pt}
\label{fig:resource_cost}
\end{figure}

Parameter reduction does not consistently reduce resource use. Fig.~\ref{fig:resource_cost} shows training times of 3.22 to 7.20 hours for insertion and 3.24 to 7.81 hours for replacement, compared with 6.60 hours for classical full fine-tuning. The classical baseline used patience 5, whereas full quantum runs used patience 2. Different stopping budgets and completed epoch counts prevent these totals from establishing a training speedup. Peak RAM usage ranges from 5.75 to 7.49~GiB for insertion and 4.92 to 8.14~GiB for replacement, compared with 6.29~GiB for the classical model.

\begin{table}[t]
\centering
\caption{Accuracy (\%) before and after attack on 40 examples.}
\vspace{-8pt}
\label{tab:robustness}
\begingroup
\scriptsize
\setlength{\tabcolsep}{1pt}
\renewcommand{\arraystretch}{0.95}
\begin{tabular*}{\columnwidth}{@{\extracolsep{\fill}}ccccccc@{}c@{}|ccccc@{}}
\hline
\multirow[c]{2}{*}{Circuit} & \multirow[c]{2}{*}{Position} & \multicolumn{5}{c}{Insertion} & & \multicolumn{5}{c}{Replacement} \\
\cline{3-7}\cline{9-13}
 & & Clean & FGSM & SPSA & HSJ & T-PGD & &
Clean & FGSM & SPSA & HSJ & T-PGD \\
\hline
\multirow[c]{4}{*}{BR4} & 1 & 72.5 & 50.0 & 65.0 & 72.5 & 20.0 & & 55.0 & 50.0 & 40.0 & 52.5 & 50.0 \\
 & 2 & 75.0 & 47.5 & 60.0 & 72.5 & 22.5 & & 37.5 & 15.0 & 12.5 & 32.5 & 32.5 \\
 & 3 & 75.0 & 47.5 & 65.0 & 75.0 & 22.5 & & 67.5 & 37.5 & 50.0 & 67.5 & 20.0 \\
 & 4 & 72.5 & 52.5 & 67.5 & 72.5 & 20.0 & & 70.0 & 50.0 & 67.5 & 70.0 & 25.0 \\
\hline
\multirow[c]{4}{*}{BR8} & 1 & 72.5 & 50.0 & 67.5 & 72.5 & 22.5 & & 55.0 & 55.0 & 55.0 & 55.0 & 55.0 \\
 & 2 & 72.5 & 52.5 & 62.5 & 72.5 & 20.0 & & 45.0 & 15.0 & 22.5 & 42.5 & 32.5 \\
 & 3 & 72.5 & 50.0 & 70.0 & 72.5 & 22.5 & & 75.0 & 40.0 & 55.0 & 72.5 & 20.0 \\
 & 4 & 77.5 & 50.0 & 65.0 & 75.0 & 22.5 & & 75.0 & 52.5 & 65.0 & 72.5 & 22.5 \\
\hline
\multirow[c]{4}{*}{SE4} & 1 & 75.0 & 50.0 & 65.0 & 72.5 & 22.5 & & 45.0 & 20.0 & 20.0 & 27.5 & 37.5 \\
 & 2 & 72.5 & 47.5 & 62.5 & 70.0 & 22.5 & & 45.0 & 20.0 & 35.0 & 40.0 & 37.5 \\
 & 3 & 72.5 & 47.5 & 62.5 & 72.5 & 20.0 & & 70.0 & 37.5 & 42.5 & 67.5 & 20.0 \\
 & 4 & 75.0 & 52.5 & 62.5 & 75.0 & 25.0 & & 72.5 & 50.0 & 62.5 & 70.0 & 22.5 \\
\hline
\multirow[c]{4}{*}{SE8} & 1 & 75.0 & 55.0 & 65.0 & 70.0 & 20.0 & & 55.0 & 55.0 & 52.5 & 55.0 & 55.0 \\
 & 2 & 75.0 & 40.0 & 55.0 & 72.5 & 17.5 & & 52.5 & 32.5 & 35.0 & 50.0 & 47.5 \\
 & 3 & 72.5 & 47.5 & 62.5 & 72.5 & 22.5 & & 72.5 & 35.0 & 52.5 & 70.0 & 20.0 \\
 & 4 & 75.0 & 50.0 & 62.5 & 75.0 & 22.5 & & 72.5 & 50.0 & 62.5 & 72.5 & 27.5 \\
\hline
\multicolumn{13}{c}{\textbf{Classical full: Clean: 72.5, FGSM: 50.0, SPSA: 65.0, HSJ: 72.5, T-PGD: 25.0}} \\
\hline
\end{tabular*}
\vspace{-15pt}
\endgroup
\end{table}
\subsection{Adversarial Robustness}

Table~\ref{tab:robustness} shows that clean-performance gains do not consistently protect against attacks. PGD-CE, PGD-margin, and the six-attack union reduce robust accuracy to 0\% for every model. Classical robust accuracy is 50.0\% under FGSM and 25.0\% under transfer PGD, whereas SE8 at q2 achieves 40.0\% and 17.5\%, respectively. Higher clean accuracy thus does not imply stronger robustness. The comparatively high HSJ results require caution given initialization failures and limited queries. Gradient diagnostics remain unresolved for 29 of the 33 models, limiting interpretation of unsuccessful searches; a failed search alone does not establish robustness.

\subsection{Discussion}
The results answer our three research questions. For RQ1, insertion benefits depend on placement and training mode, with selected configurations improving accuracy without consistent resource savings. For RQ2, the final decoder feed-forward sublayer can be replaced while preserving accuracy in this run, whereas projector and early decoder replacements perform poorly. For RQ3, neither insertion nor replacement provides consistent robustness gains under the evaluated attacks.
These findings are limited to one backbone, one dataset, and one seed. The development split supports both checkpoint selection and reporting, and the best configuration is selected from many candidates. Held-out evaluation, multiple seeds, and parameter-matched classical controls are needed to confirm gains and isolate the circuit's contribution. Unequal stopping budgets limit runtime comparisons, while the 40-example attack subset and unresolved gradient diagnostics limit robustness conclusions. The study covers image perturbations under classical simulation, not privacy attacks or quantum-hardware performance.


\section{Conclusion}
\label{sec:conclusion}
We presented a framework for evaluating quantum-layer insertion and replacement in a VLM through classical simulation. By considering predictive performance, computational cost, and adversarial robustness together, this work provides a basis for assessing when quantum integration is useful. Its value depends on where the circuit is placed, which classical component it replaces, and how the model is trained. Establishing quantum advantage requires further evidence beyond the comparisons presented here. Future work will evaluate multiple seeds and held-out data, use parameter-matched classical controls and matched training budgets, and strengthen robustness validation.

\bibliographystyle{IEEEtran}
\bibliography{References}

\end{document}